\documentclass[aps,prd,reprint,twoside,superscriptaddress,showpacs,floatfix]{revtex4-1}

\usepackage{graphicx}
\usepackage{dcolumn}
\usepackage{bm}
\usepackage{natbib}
\usepackage{multirow}
\usepackage{amsmath}
\usepackage{amssymb}
\usepackage{mathrsfs}
\usepackage{bbold}
\usepackage{hyperref}
\usepackage{aas_macros}
\usepackage{booktabs}

\newcommand{\hGpc}{{h^{-1}\mathrm{Gpc}}}

\newcommand{\hMsun}{h^{-1}\mathrm{M}_{\odot}}
\newcommand{\dg}{\boldsymbol{\delta}_\mathrm{g}}
\newcommand{\dm}{\delta}

\newcommand{\dd}{\boldsymbol{\delta}}
\newcommand{\e}{\boldsymbol{\epsilon}}
\newcommand{\E}{\boldsymbol{\mathcal{E}}}
\newcommand{\Em}{\mathcal{E}_-}
\newcommand{\Ep}{\mathcal{E}_+}
\newcommand{\Epm}{\mathcal{E}_\pm}
\newcommand{\A}{\boldsymbol{\mathcal{A}}}
\newcommand{\C}{\mathbf{C}}

\newcommand{\I}{\mathbf{I}}

\newcommand{\bg}{\boldsymbol{b}}
\newcommand{\bb}{\boldsymbol{b}_\mathrm{g}}
\newcommand{\M}{\boldsymbol{M}}
\newcommand{\Mr}{\boldsymbol{\mathcal{M}}}

\newcommand{\V}{\boldsymbol{V}}
\newcommand{\Om}{\Omega_\mathrm{m}}
\newcommand{\Ob}{\Omega_\mathrm{b}}
\newcommand{\Ok}{\Omega_\mathrm{k}}
\newcommand{\Ol}{\Omega_\Lambda}
\newcommand{\rhom}{\bar{\rho}_\mathrm{m}}

\newcommand{\T}{^{\dag}}
\newcommand{\pT}{^{\phantom{\dagger}}}

\begin{document}

\title{Optimal Weighting in Galaxy Surveys: \\ Application to Redshift-Space Distortions}

\author{Nico Hamaus} \email{hamaus@physik.uzh.ch}
\affiliation{Institute for Theoretical Physics, University of Zurich, 8057 Zurich, Switzerland}
\author{Uro{\v s} Seljak}
\affiliation{Institute for Theoretical Physics, University of Zurich, 8057 Zurich, Switzerland}
\affiliation{Physics Department, Astronomy Department and Lawrence Berkeley National Laboratory, University of California, Berkeley, California 94720, USA}
\affiliation{Ewha University, Seoul 120-750, S. Korea}
\author{Vincent Desjacques}
\affiliation{D\'epartement de Physique Th\'eorique \& Center for Astroparticle Physics, Universit\'e de Gen\`eve, 24 Quai Ernest Ansermet, 1211 Gen\`eve 4, Switzerland}

\date{\today}

\begin{abstract}
Using multiple tracers of large-scale structure allows to evade the limitations imposed by sampling variance for some parameters of interest in cosmology. We demonstrate the optimal way of carrying out a multitracer analysis in a galaxy redshift survey by considering the principal components of the shot noise matrix from two-point clustering statistics. We show how to construct two tracers that maximize the benefits of sampling variance and shot noise cancellation using optimal weights.
On the basis of high-resolution $N$-body simulations of dark matter halos we apply this technique to the analysis of redshift-space distortions and demonstrate how constraints on the growth rate of structure formation can be substantially improved. The primary limitations are nonlinear effects, which cause significant biases in the method already at scales of $k<0.1h\mathrm{Mpc}^{-1}$, suggesting the need to develop nonlinear models of redshift-space distortions in order to extract the maximum information from future redshift surveys. Nonetheless we find gains of a factor of a few in constraints on the growth rate achievable when merely the linear regime of a galaxy survey like EUCLID is considered.
\end{abstract}


\maketitle

\setcounter{footnote}{0}

\section{Introduction}
One of the deepest mysteries of contemporary cosmology is the nature of the observed accelerated expansion of the Universe. So far, its evolution can be described remarkably well by Einstein's theory of gravitation including a nonzero cosmological constant $\Lambda$. However, in order to fit the astronomical observations (e.g.,~\cite{Perlmutter1999}), $\Lambda$ must be many orders of magnitude smaller than what our standard model of particle physics would expect. This \emph{hierarchy problem} inspired various departures from the cosmological standard model, such as modifications of Einstein's field equations or the introduction of exotic forms of matter.

A particularly sensitive probe of cosmology is the large-scale structure (LSS) of the Universe. Galaxy redshift surveys map out large fractions of its observable volume and thereby reconstruct a three-dimensional map of density fluctuations whose statistical properties directly relate to fundamental cosmological parameters \cite{BigBOSS,EUCLID,SDSS-III}. Unfortunately, this reconstruction is hampered by the fact that galaxies are \emph{biased} and \emph{stochastic} tracers of the dominating dark matter density field. Even if the density field could be inferred perfectly well, the finite number $N_k$ of independent Fourier modes in the survey sets a fundamental lower limit on the achievable uncertainty, which is known as \emph{sampling variance} (or \emph{cosmic variance}, in case the survey size is the whole observable Universe). For example, in a measurement of the dark matter power spectrum $P$, the sampling variance limit is given by $\sigma_P/P\ge\sqrt{2/N_k}$, 
an uncertainty floor that propagates into all the parameters one wants to infer from $P$. This limit decreases towards smaller scales as more and more Fourier modes can be sampled, but at the same time linear theory starts to break down and higher-order perturbation theory has to be adopted to model $P$ (see, e.g., \cite{Bernardeau2002,Crocce2006a,Taruya2008a,Matsubara2008}). Further complication arises in relating the observed galaxy power spectrum to the latter, as galaxy bias becomes nonlinear and nonlocal \cite{ChuenChan2012,Baldauf2012}.

An alternative approach to accurately probe cosmological parameters is to consider multiple tracers of the density field within the well-understood linear regime. The relative clustering amplitude between multiple tracers can be inferred without sampling variance limitation because the underlying density fluctuations cancel out in taking ratios~\cite{Seljak2009a,McDonald2009}. Therefore, any cosmological information that remains in the relative clustering amplitude between different tracers can potentially be inferred with a much higher accuracy.

In this paper we focus on a particular contribution to the clustering amplitude of galaxies coming from redshift-space distortions (RSD). These are caused by peculiar velocities along the line of sight, causing their clustering statistics to become anisotropic. First treated as a contamination, this effect has been realized to be a powerful probe of cosmology, as an understanding of RSD allows to infer the growth rate of structure formation, which is directly tied to the expansion history of the Universe as well as the theory of gravity~(e.g.,  \cite{Guzzo2008,Percival2009,Cabre2009,Crocce2011,Blake2011,Ross2011,Nusser2012,Beutler2012,Reid2012,Samushia2012b,Raccanelli2012}).

After recapping the fundamentals of RSD and introducing a general formalism for the multitracer analysis in Sec.~\ref{sec1}, we present our results on the RSD analysis from $N$-body simulations in Sec.~\ref{sec2}. Finally we draw our conclusions in Sec.~\ref{sec3}.

\section{Formalism \label{sec1}}

\subsection{Galaxies in redshift space}
In galaxy surveys, radial distances are inferred via the individual redshift of objects, assuming they follow the Hubble flow. Due to gravitational attraction, however, galaxies (respectively, their host halos) build up peculiar velocities $\mathbf{v}$ which contribute to their redshift via the Doppler effect. Hence, the real-space and redshift-space locations $\mathbf{r}$ and $\mathbf{s}$ of a galaxy are related as
\begin{equation}
\mathbf{s}=\mathbf{r}+\frac{\mathbf{v}\cdot\hat{\mathbf{r}}}{H(z)}\hat{\mathbf{r}} \;, \label{RSD}
\end{equation}
where $\hat{\mathbf{r}}$ is the unit vector along the line of sight and $H(z)$ is the Hubble constant as a function of redshift $z$. On large scales, gravity causes coherent infall of test particles into the potential wells of the dark matter. Thus, galaxies are moving towards overdense regions in the Universe, resulting in an enhancement of their inferred overdensity along the line of sight. According to linear perturbation theory, the redshift-space and real-space galaxy overdensities are related as
\begin{equation}
\delta_\mathrm{g}^{(s)}(\mathbf{k},\mu)=\delta_\mathrm{g}^{(r)}(\mathbf{k})+f\mu^2\dm(\mathbf{k}) \;, \label{Kaiser}
\end{equation}
where $\mu$ the cosine of the angle between any wave vector $\mathbf{k}$ in the survey and the line of sight, $\dm$ the dark matter overdensity field and $f$ the \emph{growth rate} of structure. This well-known result \cite{Kaiser1987} further makes use of the plane-parallel approximation, assuming the separation of any galaxy pair to be much smaller than their distance to the observer. On nonlinear scales, random motions are generated in the process of virialization, which causes a damping in the clustering amplitude of galaxies along the line of sight. This so-called Finger-of-God effect is often modeled phenomenologically by an additional Gaussian damping factor in Eq.~(\ref{Kaiser})~\cite{Peacock1994}, but more elaborate schemes have been developed~(e.g., \cite{Scoccimarro2004,Tinker2006,Reid2011,Jennings2011,Sato2011,Kwan2012,Samushia2012a,DelaTorre2012,Bianchi2012}). We will neglect nonlinear corrections in this paper and focus on the large linear scales.

The growth rate $f$ is the logarithmic derivative of the growth factor $D$ with respect to the scale factor $a$. In linear theory, it can be expressed as
\begin{equation}
f\equiv d\ln D/d\ln a\simeq\Om^\gamma \;,
\end{equation}
with the matter density parameter $\Om$ and the \emph{growth index} $\gamma$ \cite{Peebles1980}. In Einstein gravity the value of $\gamma$ is about $0.55$, but can take on distinctly different values in modified gravity scenarios \cite{Linder2007}. Therefore, constraints on the growth rate can provide viable tests on the theory of gravitation.

Galaxies only form in specific, discrete locations of the density field; they are referred to as \emph{biased} and \emph{stochastic} tracers of the dark matter. On linear scales, this relation can be described locally by
\begin{equation}
\delta_\mathrm{g}^{(r)}(\mathbf{r})=b_\mathrm{g}\dm(\mathbf{r})+\epsilon \;,
\end{equation}
where the factor $b_\mathrm{g}$ is the \emph{linear galaxy bias} and $\epsilon$ a random variable denoted as \emph{shot noise}, describing the stochastic nature of this relation. Both $b_\mathrm{g}$ and $\epsilon$ depend on redshift, as well as various properties of the type of galaxies one is considering (e.g., luminosity, color, host-halo mass).

Together with Eq.~(\ref{Kaiser}) this yields a model for the overdensity field of galaxies in redshift space,
\begin{equation}
\delta_\mathrm{g}^{(s)}(\mathbf{k},\mu)=\left(b_\mathrm{g}+f\mu^2\right)\dm(\mathbf{k})+\epsilon \;. \label{Model}
\end{equation}
Since the phenomena of galaxy biasing and RSD are to multiply the density field $\dm$ by some factor, we can simply define a more general \emph{effective bias} parameter $b$ that contains both contributions,
\begin{equation}
b\equiv b_\mathrm{g}+f\mu^2 \;. \label{b_eff}
\end{equation}
In the following, we will drop the superscript that distinguishes between real-space and redshift-space quantities for clarity. If not explicitly mentioned otherwise, all symbols should be understood as given in redshift space.

\subsection{Multiple tracers}
In order to exploit the gains of sampling variance and shot noise cancellation, we need to consider multiple tracers of the dark matter density field \cite{Seljak2009a,McDonald2009}. One way to achieve this is splitting some galaxy catalog into bins of a certain observable property of the galaxies (like luminosity, color, host-halo mass, etc.). However, the following framework is not limited to galaxies and may be adopted for other tracers of the dark matter density field as well.

\subsubsection{Covariance matrix}
We start by writing the density fields of $N$ tracers as a vector $\dg \equiv \left(\delta_{\mathrm{g}_1},\delta_{\mathrm{g}_2},\dots,\delta_{\mathrm{g}_N}\right)$. The outer product of this vector, once ensemble averaged within a $k$-shell in Fourier space, yields the covariance matrix $\C\equiv\langle\dg\pT\dg\T\rangle$, where the $\T$ symbol denotes the Hermitian conjugate (transpose and complex conjugation). Plugging in the model for galaxy overdensities in redshift space from Eq.~(\ref{Model}), it reads
\begin{equation}
\C=\bg\bg\T P + \E \;, \label{Cov}
\end{equation}
with the effective bias vector $\bg=\bb+f\mu^2\I$, the dark matter power spectrum $P\equiv\left<\dm\dm^*\right>$ and the \emph{shot noise matrix} $\E\equiv\langle\e\e\T\rangle$ (by definition $\langle\e\dm^*\rangle=0$). The covariance matrix contains all auto-power and cross-power spectra of the considered tracers, so in total $N(N-1)/2$ independent elements per $k$-shell. However, since all tracers follow the same dark matter density distribution, these elements are correlated.

Let us compare the analysis for a single tracer with the one for two tracers. From a single tracer we can only observe an estimator of its auto-power spectrum
\begin{equation}
C(k,\mu)=\left(1+\beta\mu^2\right)^2b_\mathrm{g}^2P(k)+\mathcal{E} \;, \label{single_tracer}
\end{equation}
where we define $\beta\equiv f/b_\mathrm{g}$. In this parametrization it is obvious to see that $b_\mathrm{g}$ is degenerate with the power spectrum $P$; respectively, its normalization $\sigma_8$ defined via
\begin{equation}
P\equiv\sigma_8^2P_0 \; , \label{sigma_8} 
\end{equation}
where $P_0$ describes the shape of the power spectrum and is assumed to be known, at least up to linear order. From Eq.~(\ref{single_tracer}), only the combination $b_\mathrm{g}^2P$ can be determined at $\mu=0$ (usually, the shot noise is assumed to be Poissonian, meaning it is scale independent and given by the inverse number density of galaxies $\bar{n}^{-1}$). The same is true for the growth rate $f$, we can only determine the product
\begin{equation}
f^2P=\frac{C(k,\mu)-\mathcal{E}-\left[C(k,0)-\mathcal{E}\right]\left(1+2\beta\mu^2\right)}{\mu^4} \;, \label{f^2P}
\end{equation}
but this does not apply for $\beta$, which can be extracted directly from observations of $C$ where $\mu\neq0$ (e.g., \cite{Okumura2011}). Note, however, that in this case the achievable error on $\beta$ is limited by the sampling variance inherent to $P$. 

In case two distinct tracers of the density field with biases $b_\mathrm{g}$ and $\alpha b_\mathrm{g}$ are observed, where $\alpha$ is their relative galaxy bias, we obtain the three following power spectra:
\begin{gather}
C_{11}(k,\mu)=\left(1+\beta\mu^2\right)^2b_\mathrm{g}^2P(k)+\mathcal{E}_{11} \;, \\
C_{22}(k,\mu)=\left(\alpha+\beta\mu^2\right)^2b_\mathrm{g}^2P(k)+\mathcal{E}_{22} \;, \\
C_{12}(k,\mu)=\left(1+\beta\mu^2\right)\left(\alpha+\beta\mu^2\right)b_\mathrm{g}^2P(k)+\mathcal{E}_{12} \;.
\end{gather}
The degree to how well they are correlated is quantified by the cross-correlation coefficient $r^2\equiv C_{12}^2/C_{11}C_{22}$, so in the idealistic case of no shot noise ($\E=0$), $r=1$ and the ratios
\begin{equation}
C_{12}/C_{11}=C_{22}/C_{12}=\sqrt{C_{22}/C_{11}}=\frac{\alpha+\beta\mu^2}{1+\beta\mu^2} \label{ratios}
\end{equation}
all yield the same expression, which is independent of $P$. Hence, a combination of observations at different $\mu$ yields $\alpha$ and $\beta$ without sampling variance \cite{McDonald2009}. Unfortunately, in realistic surveys $\E\neq0$, so $P$ will not cancel out completely and the three ratios will be different and scale dependent. This gives rise to residual sampling variance inherent to a measurement of $\alpha$ and $\beta$, which in general can be much smaller than in the single-tracer case. In turn, a better measurement of $\beta$ allows a more precise estimate on $f^2P$ [see Eq.~(\ref{f^2P})]. However, the accuracy on $f^2P$ is still limited by sampling variance, yielding $\sigma_{f^2P}/f^2P\ge\sqrt{2/N_k}$, respectively $\sigma_{f\sigma_8}/f\sigma_8\ge\sqrt{1/2N_k}$~\cite{McDonald2009}.

If both $\dg$ \emph{and} $\dm$ are known, we can simply add the dark matter overdensity mode $\dm$ to the overdensities of the tracers and write $\dd \equiv \left(\dm,\dg\right)$. In this case, the effective bias can be obtained directly by taking the ratio $\langle\dg\dm^*\rangle\big/\langle\dm\dm^*\rangle$. The covariance matrix then becomes
\begin{equation}
\C = \left( \begin{array}{cc}
\langle\dm\dm^*\rangle & \langle\dm\dg\T\rangle \vspace{1pt} \\
\langle\dg\dm^*\rangle & \langle\dg\pT\dg\T\rangle \\
\end{array} \right) =
\left( \begin{array}{cc}
P & \bg\T P \\
\bg P & \bg\bg\T P + \E \\
\end{array} \right) \;, \label{Covm}
\end{equation}
with $N(N+1)/2$ independent elements. Now the degeneracy between $b_\mathrm{g}$, $\sigma_8$ and $f$ is lifted because $P$ is known separately. Considering the cross-correlation coefficient of a tracer $\delta_\mathrm{g}$ with the dark matter $\delta$, we find
\begin{equation}
r^2=\frac{\langle\delta_\mathrm{g}\dm^*\rangle^2}{\langle\delta_\mathrm{g}^{\phantom{*}}\delta_\mathrm{g}^*\rangle\langle\dm\dm^*\rangle}=\left(1+\frac{\mathcal{E}}{\left(b_\mathrm{g}+f\mu^2\right)^2P}\right)^{-1} \;,
\end{equation}
so the deviation of $r$ from unity is only controlled by the shot noise of the tracer. Again, if $\mathcal{E}=0$, the dark matter power spectrum disappears completely, so that $b_\mathrm{g}$ and $f$ can be determined without sampling variance.

\subsubsection{Fisher information}
In order to determine more quantitatively how much information on cosmology is buried in the clustering statistics of biased tracers and the dark matter, we have to compute the Fisher information matrix \cite{Fisher1935}, a derivation of which is presented in the following: we start with a multivariate Gaussian likelihood of the data vector $\dg$
\begin{equation}
\mathscr{L}=\frac{1}{(2\pi)^{N/2}\sqrt{\det\C}}\exp\left(-\frac{1}{2}\dg\T\C^{-1}\dg\pT\right) \;, \label{likelihood}
\end{equation}
which is a reasonable assumption on large scales, where $\dm\ll1$. The Fisher information matrix for the parameters $\theta_i$ and $\theta_j$ is obtained by ensemble averaging over the Hessian of the log-likelihood \cite{Tegmark1997,Heavens2009},
\begin{equation}
F_{ij} \equiv -\left<\frac{\partial^2\ln\mathscr{L}}{\partial\theta_i\partial\theta_j}\right>=\frac{1}{2}\mathrm{Tr}\left(\frac{\partial\C}{\partial\theta_i}\C^{-1}\frac{\partial\C}{\partial\theta_j}\C^{-1}\right) \;. \label{fisher}
\end{equation}
According to the model from Eq.~(\ref{Cov}), the derivative of the halo covariance matrix with respect to the parameters is
\begin{equation}
\frac{\partial\C}{\partial \theta_i} = \left(\bg\bg_i\T+\bg_i\bg\T\right)P+\bg\bg\T P_i \;, \label{C'}
\end{equation}
where $\bg_i\equiv\partial\bg/\partial\theta_i$, $P_i\equiv\partial P/\partial\theta_i$ and we assume $\partial\E/\partial\theta_i=0$. Utilizing the \emph{Sherman-Morrison} formula \cite{Sherman1950,Bartlett1951}, the inverse of the covariance matrix becomes
\begin{equation}
\C^{-1}=\E^{-1}-\frac{\E^{-1}\bg\bg\T\E^{-1}P}{1+\bg\T\E^{-1}\bg P} \;, \label{Sherman-Morrison}
\end{equation}
provided the shot noise matrix is not singular (e.g., with vanishing $\E$ also $\C$ becomes singular by construction). The full Fisher matrix for this case is calculated in Appendix~\ref{app_A}. The result is
\begin{multline}
F_{ij}=\left[\Sigma^{-1}\left(\Sigma_{ij}+\frac{\Sigma_i\Sigma_j}{\Sigma}\right) +
\left(\Sigma_{ij}-\frac{\Sigma_i\Sigma_j}{\Sigma}\right)\right. \\
\left.+\frac{\Sigma_i}{\Sigma}\frac{P_j}{P} + 
\frac{\Sigma_j}{\Sigma}\frac{P_i}{P} +
\frac{P_iP_j}{2P^2}\right]\left(1+\Sigma^{-1}\right)^{-2}\;, \label{F_h}
\end{multline}
where
\begin{equation}
\Sigma_{ij}\equiv\bg_i\T\E^{-1}\bg_j P \;,
\end{equation}
and $\Sigma_i$ as well as $\Sigma$ are defined accordingly by simply omitting the corresponding indices (derivatives). We can identify the first two terms in the square brackets as a \emph{single-tracer} and a \emph{multitracer} term, respectively. The single-tracer term is suppressed by a factor $\Sigma^{-1}$ as compared to the multitracer term. By definition, the latter vanishes for the case of only one tracer, since then $\Sigma\Sigma_{ij}=\Sigma_i\Sigma_j$ and Eq.~(\ref{F_h}) simplifies to
\begin{equation}
F_{ij}=\left(2\frac{b_ib_j}{b^2}+\frac{b_i}{b}\frac{P_j}{P}+\frac{b_j}{b}\frac{P_i}{P}+\frac{P_iP_j}{2P^2}\right)\left(1+\frac{\mathcal{E}}{b^2P}\right)^{-2}\;. \label{F1}
\end{equation}
While the Fisher information for multiple tracers can in principle become infinite in the limit of no shot noise, Eq.~(\ref{F1}) reaches a finite limit when $\mathcal{E}\rightarrow0$. 

We consider two sets of parameters separately: the ones that influence only the effective bias,
\begin{equation}
\boldsymbol{\theta}^{(b)}\equiv\left(\bb,f\right)
\end{equation}
and the ones that go into the matter power spectrum,
\begin{equation}
\boldsymbol{\theta}^{(P)}\equiv\left(\sigma_8,n_s,h,\Ol,\Om,\Ob,\Ok\right) \;.
\end{equation}
The elements of the shot noise matrix are usually not considered as quantities of interest, but as nuisance parameters that can be marginalized over. From Eq.~(\ref{F_h}) it is evident that there are degeneracies between $\boldsymbol{\theta}^{(b)}$ and $\boldsymbol{\theta}^{(P)}$ due to mixed terms of the form $\Sigma_iP_j$. However, if $\Sigma$ is sufficiently large, those terms are suppressed in comparison to the term $\Sigma_{ij}$.

In case the dark matter density field is known in addition to the galaxies, the derivative of $\C$ from Eq.~(\ref{Covm}) becomes
\begin{equation}
\frac{\partial\C}{\partial\theta_i} = \left( \begin{array}{cc}
P_i & \bg_i\T P + \bg\T P_i \\
\bg_iP + \bg P_i \;\; &\;\; \bg\bg_i\T P+\bg_i\bg\T P+\bg\bg\T P_i \\
\end{array} \right) \;. \label{Covm'}
\end{equation}
Furthermore, a block inversion of $\C$ yields
\begin{equation}
\C^{-1} = \left( \begin{array}{cc}
(1+\Sigma)P^{-1} & -\bg\T\E^{-1} \\
-\E^{-1}\bg & \E^{-1} \\
\end{array} \right) \;, \label{CovmI}
\end{equation}
and the Fisher information becomes (see Appendix~\ref{app_A})
\begin{equation}
F_{ij}=\Sigma_{ij} + \frac{P_iP_j}{2P^2} \;. \label{F_m}
\end{equation}
For a single tracer this further simplifies to
\begin{equation}
F_{ij}=\frac{b_ib_j P}{\mathcal{E}} + \frac{P_iP_j}{2P^2} \;. \label{F_m1}
\end{equation}
This expression may increase indefinitely for sufficiently small $\mathcal{E}$. In the limit $\mathcal{E}\rightarrow0$, the effective bias can be determined exactly, allowing an exact measurement of the parameters $\boldsymbol{\theta}^{(b)}$~\cite{McDonald2009,Seljak2009b}. On the other hand, constraints on the parameters $\boldsymbol{\theta}^{(P)}$ are always limited by the variance of the power spectrum $\mathrm{Var}(P)=2P^2$. Here, there are no mixed terms depending both on the effective bias and the power spectrum, so parameter degeneracies between $\boldsymbol{\theta}^{(b)}$ and $\boldsymbol{\theta}^{(P)}$ are absent. We note that in the limit of small $\Sigma_i/\Sigma$, Eq.~(\ref{F_h}) reduces to Eq.~(\ref{F_m}).

\subsubsection{Reparametrization}
As mentioned above, the growth rate $f$ cannot be determined from a galaxy redshift survey alone, since it is degenerate with the power spectrum $P$, respectively its normalization $\sigma_8$. This degeneracy can only be broken with knowledge of the dark matter density field, or other prior constraints. For this reason it is sometimes convenient to reparametrize by the mapping
\begin{gather}
\tilde{\boldsymbol{\theta}}^{(b)}=\boldsymbol{\theta}^{(b)}\!\!\left/b^{\mathrm{ref}}_\mathrm{g}\right.=\left(\bb,f\right)\!\left/b^{\mathrm{ref}}_\mathrm{g}\right.\equiv\left(\boldsymbol{\alpha},\beta\right)\;,\\
\tilde{\sigma}_8=\sigma_8b^{\mathrm{ref}}_\mathrm{g}\;,
\end{gather}
where $b^{\mathrm{ref}}_\mathrm{g}$ is an arbitrary reference galaxy bias so that $\boldsymbol{\alpha}$ is the relative galaxy bias of all considered tracers to this reference. This mapping leaves the covariance matrix from Eq.~(\ref{Cov}) unchanged. We can conveniently choose the lowest bias of all tracers as the reference, $b^{\mathrm{ref}}_\mathrm{g}=b_{\mathrm{g}_1}$, such that $\boldsymbol{\alpha}=\left(1,b_{\mathrm{g}_2}/b_{\mathrm{g}_1},\dots,b_{\mathrm{g}_N}/b_{\mathrm{g}_1}\right)$ is always larger than unity and $\beta=f/b_{\mathrm{g}_1}$. Another popular parametrization~is
\begin{gather}
\tilde{\boldsymbol{\theta}}^{(b)}=\boldsymbol{\theta}^{(b)}\sigma_8^{\mathrm{ref}}=\left(\bb\sigma_8^{\mathrm{ref}},f\sigma_8^{\mathrm{ref}}\right)\;,\\
\tilde{\sigma}_8=\sigma_8/\sigma_8^{\mathrm{ref}} \;,
\end{gather}
with $\sigma_8^{\mathrm{ref}}$ being some arbitrary reference normalization of the power spectrum, which we choose to be identical with our simulation input value of $\sigma_8=0.81$. In this paper we will quote constraints on both $\beta$ and $f\sigma_8$ when considering a galaxy redshift survey only, and on $f$ when the latter is combined with dark matter observations.

\section{Analysis \label{sec2}}

\subsection{Numerical setup}
Our numerical analysis is based on a high-resolution $N$-body simulation performed at the University of Z{\"u}rich supercomputer zBox3 with the {\scshape gadget}-2 code \cite{Springel2005b}. It contains $1536^3$ particles of mass $4.7\times10^{10}\hMsun$ in a box of $1.3\hGpc$ a side. We chose our fiducial cosmology to match the WMAP5 best fit with $\sigma_8=0.81$, $n_s=0.96$, $h=0.7$, $\Ol=0.721$, $\Om=0.279$, $\Ob=0.046$, $\Ok=0$ \cite{Komatsu2009}. We further employ a friends-of-friends algorithm \cite{Davis1985} with a linking length of $20\%$ of the mean interparticle distance to generate halo catalogs at different redshifts. With a minimum of $20$ dark matter particles per halo we resolve halo masses down to $M_{\mathrm{min}}\simeq 9.4\times10^{11}\hMsun$, resulting in a mean halo number density of $\bar{n}\simeq4.0\times10^{-3}h^3\mathrm{Mpc}^{-3}$ at $z=0$ and $\bar{n}\simeq3.3\times10^{-3}h^3\mathrm{Mpc}^{-3}$ at $z=1$.

In order to transform the real-space halo catalog into redshift space, we apply Eq.~(\ref{RSD}) using the velocities of the halos along the three independent directions of the box ($x$, $y$ and $z$ axis). Thus, three independent redshift-space catalogs can be constructed from a single real-space catalog, yielding a total effective volume of $V_\mathrm{eff}\simeq6.6h^{-3}\mathrm{Gpc}^3$. Density fields are created by cloud-in-cell interpolation \cite{Hockney1988} onto a mesh of $1024^3$ grid points, and the Fourier modes are obtained using a FFT algorithm.

We utilize the {\scshape idl} algorithm {\scshape mpfit} \cite{Markwardt2009} to fit our models to the numerical data and find the best-fit parameters including their uncertainties. It is based on the {\scshape minpack} distribution by \cite{More1978} and uses the Levenberg-Marquardt technique to find the minimum of a multidimensional nonlinear least-squares problem. Parameter uncertainties are calculated via the Jacobian of the chi square, which is determined numerically using finite-difference derivatives.

\subsection{Optimal tracer selection}
In order to fully exploit the benefits of the multitracer approach, the question remains on how to ideally construct different tracers from a given galaxy catalog. In this section we will derive the answer to that question and test it on the basis of $N$-body simulations.

\begin{figure}[!t]
\centering
\resizebox{\hsize}{!}{
\includegraphics[trim=0 0 0 0,clip]{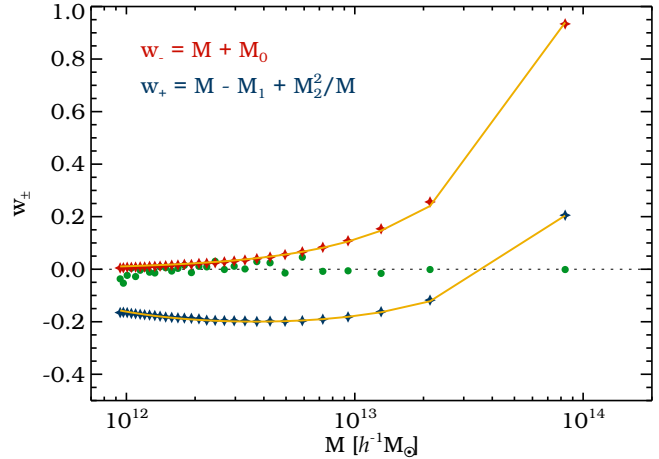}}
\caption{Three different eigenvectors of the shot noise matrix obtained from a halo catalog that has been split into $30$ mass bins of equal number density. The two non-Poisson eigenvectors (upper red and lower blue stars) are overplotted with their best-fit weighting functions $w_\pm$ as solid lines with functional form given in the top left of the panel in corresponding colors. One representative Poisson eigenvector is shown as green circles; when used as a weighting function it yields a very low clustering signal-to-noise ratio $\Sigma$, due to its oscillatory behavior.}
\label{fig1}
\end{figure}

\subsubsection{Principal components of the signal-to-noise ratio}
A quantity that plays a crucial role for the cosmological information content contained in the two-point clustering statistics of LSS is the clustering signal-to-noise ratio $\Sigma\equiv\bg\T\E^{-1}\bg P$. For dark matter halos, it has been demonstrated on the basis of numerical $N$-body simulations that $\Sigma$ is dominated by only two \emph{principal components} corresponding to the two nontrivial (non-Poisson) eigenvectors of the shot noise matrix $\E$ (see Fig.~6 in \cite{Hamaus2010}). We denote these components with a plus and a minus subscript, according to their super- and sub-Poissonian shot noise levels, respectively. In this manner we can expand the clustering signal-to-noise ratio as
\begin{equation}
\Sigma\equiv\bg\T\E^{-1}\bg P=\sum\limits_{i=0}^N\frac{\left(\V^\intercal_i\bg\right)^2}{\lambda_i}P\simeq\frac{b_-^2}{\Em}P+\frac{b_+^2}{\Ep}P \;, \label{S_N}
\end{equation}
where $\V^\intercal_i$ and $\lambda_i$ are the $N$ eigenvectors and eigenvalues of the shot noise matrix $\E$, and $b_\pm$ and $\Epm$ are the effective bias and shot noise of the two principal components of the tracer density field,
\begin{equation}
\delta_\pm\equiv\frac{\sum_i w_\pm(M_i)\delta_{\mathrm{g}}(M_i)}{\sum_i w_\pm(M_i)} \;,
\end{equation}
where the summation runs over all individual objects in the volume. The $w_\pm(M)$ are weighting functions corresponding to the two non-Poisson eigenvectors of the shot noise matrix, as depicted in Fig.~\ref{fig1}. Equation~(\ref{S_N}) states that a splitting of the tracer density field into $N$ mass bins yields, in the limit $N\rightarrow\infty$, about the same clustering signal-to-noise ratio as simply considering the two weighted fields $\delta_+$ and $\delta_-$. Therefore, these are the most promising candidates to carry out a multitracer analysis with.

In \cite{Hamaus2010} the functional form of $w_-(M)$ was found to be well described by
\begin{equation}
w_-(M)=M+M_0 \;,
\end{equation}
where $M_0$ is about three times the minimum halo mass resolved in the simulation, $M_0\simeq3M_\mathrm{min}$. For the second weighting function we find
\begin{equation}
w_+(M)=M-M_1+M_2^2/M
\end{equation}
to be a good fit to the second eigenvector of the shot noise matrix, as shown in Fig.~\ref{fig1}. For the constants $M_1$ and $M_2$ we find $M_1\simeq1\times 10^{14}\hMsun$, $M_2\simeq3\times 10^{13}\hMsun$ at $z=0$ and $M_1\simeq1\times 10^{14}\hMsun$, $M_2\simeq1\times 10^{15}\hMsun$ at $z=1$.

The physical origin of the first principal component is related to halo exclusion effects \cite{Smith2007}. The sampling of the density field with halos is less stochastic than Poisson sampling (sampling with points) due their finite extension. Since the exclusion volume of halos is proportional to their mass, this effect is strongest at the high mass end \cite{Hamaus2010}. The second principal component can be interpreted as a loop correction to the galaxy bias \cite{Hamaus2011}, coming from the second-order term in a local bias expansion model \cite{Fry1993}. Due to its nonlinear character, it adds a super-Poissonian shot noise contribution to the two-point clustering statistics of halos, originating from the squared density field \cite{Heavens1998,McDonald2006}. However, through mode coupling it also yields a second-order clustering signal that originates from the bispectrum (three-point function) and adds valuable information coming from smaller scales.

\begin{figure}[!t]
\centering
\resizebox{\hsize}{!}{
\includegraphics[trim=0 0 0 0,clip]{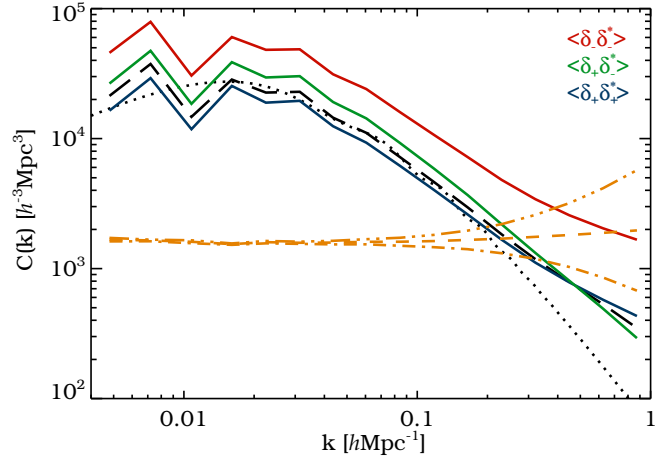}}
\caption{Real-space auto-power and cross-power spectra of the two fields $\delta_+$ and $\delta_-$ (as indicated), obtained through weighting the halo catalog by the two non-Poisson eigenvectors of the shot noise matrix $w_+$ and $w_-$. The long-dashed black line shows the dark matter power spectrum from the simulation and the dotted line its linear theory prediction. The three different estimators for the relative bias $\alpha$ between $\delta_+$ and $\delta_-$, as defined in Eq.~(\ref{ratios}), are depicted in dot-dashed ($C_{+-}/C_{++}$), dot-dot-dot-dashed ($C_{--}/C_{+-}$) and dashed ($\sqrt{C_{--}/C_{++}}$). For visibility, they were shifted upwards by a factor of $10^3$.}
\label{fig2}
\end{figure}

Figure \ref{fig2} displays the three covariance matrix elements $C_{++}=\left<\delta_+^{\phantom{*}}\delta_+^*\right>$, $C_{+-}=\left<\delta_+^{\phantom{*}}\delta_-^*\right>$ and $C_{--}=\left<\delta_-^{\phantom{*}}\delta_-^*\right>$ in real space ($\mu=0$) at redshift $z=0$, extracted from our simulation. The three power spectra are obviously highly correlated and closely follow the shape of the estimated dark matter power spectrum up to $k\simeq0.1h\mathrm{Mpc}^{-1}$. By construction, the shot noises of the two fields are not correlated, i.e., $\langle\epsilon_-\epsilon_+\rangle=0$. Taking the ratios as in Eq.~(\ref{ratios}) yields three possible estimators for the relative galaxy bias $\alpha\equiv b_\mathrm{g_-}/b_\mathrm{g_+}\simeq1.6$. In the following two subsections, we will provide evidence for the claim that the fields $\delta_+$ and $\delta_-$ are indeed the optimal choice for a multitracer analysis.

\subsubsection{Sampling variance cancellation}
The idea of utilizing multiple tracers is to cancel sampling variance from the underlying density field $\dm$. To quantify the magnitude of cancellation between any two tracers $\delta_{\mathrm{g}_1}$ and $\delta_{\mathrm{g}_2}$, we define the following statistic:
\begin{equation}
\sigma^2_\mathrm{SV}\equiv\frac{\langle\left|b_2\delta_{\mathrm{g}_1}(\mathbf{k},\mu)-b_1\delta_{\mathrm{g}_2}(\mathbf{k},\mu)\right|^2\rangle}{\langle\left|b_2\delta_{\mathrm{g}_1}(\mathbf{k},\mu)\right|^2\rangle+\langle\left|b_1\delta_{\mathrm{g}_2}(\mathbf{k},\mu)\right|^2\rangle} \;.
\end{equation}
Here, $b_1$ and $b_2$ is the effective bias as defined in Eq.~(\ref{b_eff}). If the two tracers are completely uncorrelated ($r=0$), there is no cancellation and $\sigma_\mathrm{SV}=1$. Yet, if they are perfectly correlated ($r=1$), $\sigma_\mathrm{SV}=0$. Because the real and imaginary parts of any given Fourier mode are uncorrelated, we can swap them for one of the tracers to mimic the case of no correlation.

\begin{table*}[!t]
\centering
\caption{Details of the tracer selection.}
\begin{ruledtabular}
\begin{tabular}{lccccccccccc}
\toprule
Selection   & $M_\mathrm{cut}$ & $\bar{M}_1$ & $\bar{M}_2$ & $\bar{n}_1$              & $\bar{n}_2$              & $\mathcal{E}_{11}$       & $\mathcal{E}_{22}$       & $\mathcal{E}_{12}$       & $b_{\mathrm{g}_1}$ & $b_{\mathrm{g}_2}$ & $\Sigma_\mathrm{max}$ \\
            & $[\hMsun]$       & $[\hMsun]$  & $[\hMsun]$  & $[h^3\mathrm{Mpc}^{-3}]$ & $[h^3\mathrm{Mpc}^{-3}]$ & $[h^{-3}\mathrm{Mpc}^3]$ & $[h^{-3}\mathrm{Mpc}^3]$ & $[h^{-3}\mathrm{Mpc}^3]$ &                    &                    &           \\
\midrule\midrule\hline\midrule\midrule
$20/80$     & $1.20\times10^{12}$ & $1.04\times10^{12}$ & $7.72\times10^{12}$ & $8.50\times10^{-4}$ & $3.15\times10^{-3}$ & 1432 & 964  & 448 & 0.899 & 0.988 & 74   \\
$50/50$     & $2.00\times10^{12}$ & $1.33\times10^{12}$ & $1.13\times10^{13}$ & $2.00\times10^{-3}$ & $2.00\times10^{-3}$ & 911  & 1243 & 547 & 0.896 & 1.042 & 73   \\
$80/20$     & $5.00\times10^{12}$ & $1.97\times10^{12}$ & $2.22\times10^{13}$ & $3.15\times10^{-3}$ & $8.50\times10^{-4}$ & 854  & 1770 & 606 & 0.901 & 1.217 & 73 \\
$u/w_-$     & -                   & $6.30\times10^{12}$ & $6.76\times10^{13}$ & $4.00\times10^{-3}$ & $4.00\times10^{-3}$ & 812  & 51   & 53  & 0.969 & 1.457 & 2067 \\
$w_+/w_-$   & -                   & $2.29\times10^{12}$ & $6.76\times10^{13}$ & $4.00\times10^{-3}$ & $4.00\times10^{-3}$ & 804  & 51   & 7   & 0.902 & 1.457 & 2098 \\
\bottomrule
\end{tabular}
\end{ruledtabular}
\label{tab1}
\end{table*}

We consider the following selection criteria for two tracers from our halo catalog:
\begin{description}
 \item[$20/80$] lightest $20\%$ vs heaviest $80\%$ of all halos,
 \item[$50/50$] lightest $50\%$ vs heaviest $50\%$ of all halos,
 \item[$80/20$] lightest $80\%$ vs heaviest $20\%$ of all halos,
 \item[$u/w_-$] all uniformly weighted vs all $w_-$-weighted halos,
 \item[$w_+/w_-$] all $w_+$-weighted vs all $w_-$-weighted halos.
\end{description}
The first three are simply obtained via cutting the halo catalog in two at different mass thresholds $M_\mathrm{cut}$ to yield the indicated abundances in each bin. The fourth selection utilizes the whole uniform halo catalog (not weighted) and its $w_-$-weighted form, while the last one uses both weighting functions $w_\pm$ to construct two tracers from one and the same halo catalog. Further details about the tracers with these selection criteria can be found in Table~\ref{tab1}.

\begin{figure}[!t]
\centering
\resizebox{\hsize}{!}{
\includegraphics[trim=0 0 0 0,clip]{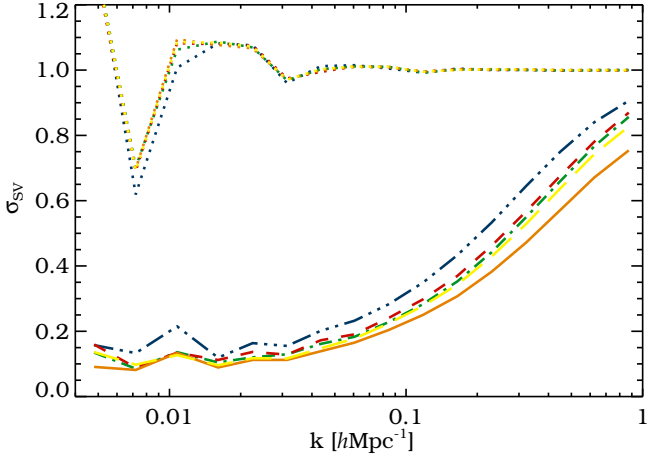}}
\caption{Sampling variance statistic $\sigma_\mathrm{SV}$ for five different tracer selections at $z=0$ as described in the text, $20/80$ (dot-dot-dot-dashed, blue), $50/50$ (dot-dashed, green), $80/20$ (dashed, red), $u/w_-$ (solid, orange) and $w_+/w_-$ (long-dashed, yellow). Dotted lines show the corresponding results for uncorrelated modes by swapping the real and imaginary parts of one of the tracer's Fourier modes.}
\label{fig3}
\end{figure}

Figure~\ref{fig3} shows $\sigma_\mathrm{SV}$ for the different tracer pairs; on large scales they all exhibit significant cancellation of sampling variance as compared to the reference case with switched real and imaginary parts (dotted lines). Towards smaller scales this effect is deteriorated due to the onset of nonlinearities and velocity dispersion (Finger-of-God effects) \cite{Scoccimarro2004}. As evident from the plot, a combination of the uniform with the $w_-$-weighted halos yields the highest cancellation of sampling variance. However, the combination of $w_+$-weighted and $w_-$-weighted halos shows a comparable suppression, as opposed to cutting the halo catalog in two, which yields less cancellation, especially for a low mass cut. If a mass cut is imposed to construct two tracers, the highest sampling variance cancellation is achieved when the same abundance of objects in each of the resulting catalogs is chosen ($50/50$), which is in agreement with the findings of \cite{Gil-Marin2010}.

\subsubsection{Fisher information}
Cancellation of sampling variance alone is not a sufficient indicator on how well cosmological parameters can be constrained in a multitracer analysis. This is because one is looking for relative changes in the clustering signal from multiple tracers, and not for the absolute clustering amplitude in each tracer. In this paragraph we will show that it is desirable to have tracers with a high relative galaxy bias ratio $\alpha$.

\begin{figure*}[!t]
\centering
\resizebox{\hsize}{!}{
\includegraphics[trim=0 0 0 0,clip]{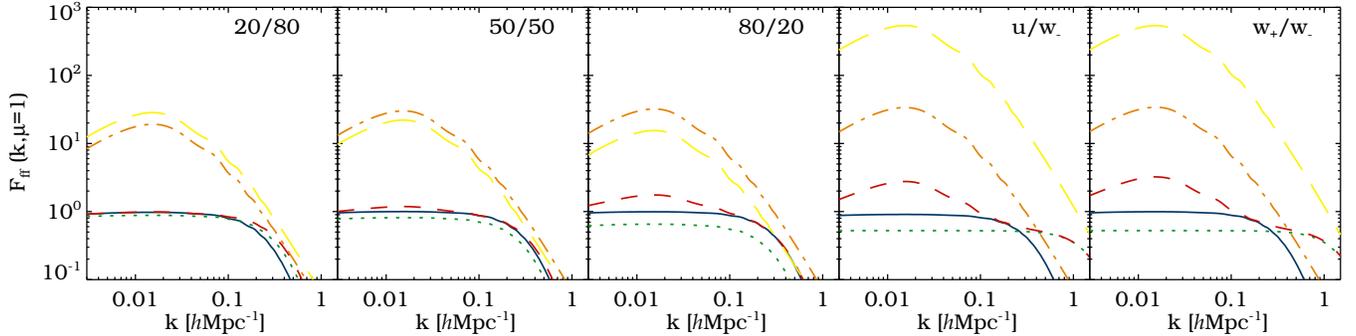}}
\caption{Fisher matrix element $F_{ff}$ as a function of $k$ at $\mu=1$ and $z=0$ for various tracer selections (indicated in the top right of each panel). Results are shown for a single-tracer analysis with the first (solid, blue) and the second tracer (dotted, green) used separately, both of the tracers combined in a multitracer analysis (dashed, red), and each of the tracers combined with idealistic dark matter observations (dot-dashed orange and long-dashed yellow, respectively).}
\label{fig4}
\end{figure*}

Quantitatively, the achievable accuracy on a given cosmological parameter $\theta$ is determined by the inversion of the Fisher matrix of Eq.~(\ref{F_h}). For the sake of simplicity, let us consider only the $f$-$f$ element,
\begin{equation}
F_{ff}=\frac{\Sigma\Sigma_{ff}+\Sigma_f^2+\Sigma\left(\Sigma\Sigma_{ff}-\Sigma_f^2\right)}{1+\Sigma^2}\;. \label{F_ff}
\end{equation}
In the high signal-to-noise regime, $\Sigma\gg1$, and expressing $\Sigma$, $\Sigma_f$, as well as $\Sigma_{ff}$ in terms of the two principal components as in Eq.~(\ref{S_N}), we get
\begin{equation}
F_{ff}\simeq\frac{2\left(\frac{b_-^2}{\mathcal{E}_-^2}+\frac{b_+^2}{\mathcal{E}_+^2}\right)+\frac{\left(b_-+b_+\right)^2}{\mathcal{E}_-\mathcal{E}_+}}{\left(\frac{b_-^2}{\mathcal{E}_-}+\frac{b_+^2}{\mathcal{E}_+}\right)^2}\mu^4 + \frac{\frac{\left(b_--b_+\right)^2}{\mathcal{E}_-\mathcal{E}_+}}{\frac{b_-^2}{\mathcal{E}_-}+\frac{b_+^2}{\mathcal{E}_+}}\mu^4P \;.
\end{equation}
As a second approximation, we can make use of the fact that $b_-^2/\mathcal{E}_-\gg b_+^2/\mathcal{E}_+$ and $\mathcal{E}_-\ll\mathcal{E}_+$ (see Table~\ref{tab1}), which yields
\begin{gather}
F_{ff}\simeq\frac{2\mu^4}{b_-^2} + \left(1-\frac{b_+}{b_-}\right)^2\frac{\mu^4P}{\mathcal{E}_+}= \nonumber \\
=\frac{2\mu^4}{(b_\mathrm{g_-}+f\mu^2)^2} + \left(1-\frac{1+\beta\mu^2}{\alpha+\beta\mu^2}\right)^2\frac{\mu^4P}{\mathcal{E}_+}
\;. \label{F_ff_exp}
\end{gather}
The single-tracer term (as derived in \cite{White2009}) is dominated by the first principal component of $\Sigma$, while the multitracer term depends on the bias ratio $\alpha=b_\mathrm{g_-}/b_\mathrm{g_+}$ of both principal components and the shot noise $\mathcal{E}_+$ of the second principle component. In order to maximize $F_{ff}$, it is thus desirable to have a large $\alpha$ and a low $\mathcal{E}_+$ at the same time. In the special case of uniform Poisson shot noise, Eq.~(\ref{F_ff}) reproduces the expression derived in \cite{Bernstein2011}.

If the dark matter density field is known separately, this Fisher matrix element becomes
\begin{equation}
F_{ff}=\Sigma_{ff}\simeq\frac{\mu^4P}{\mathcal{E}_-} \;, \label{F_ff_m}
\end{equation}
thus independent of the effective bias and limited only by the low shot noise level of the first principal component.

Figure \ref{fig4} depicts $F_{ff}(k,\mu=1)$ for all of our five tracer selections. For each case we further distinguish between the following scenarios:
\begin{itemize}
\item single-tracer analysis with each of the two tracers taken separately,
\item multiple-tracer analysis with both tracers combined,
\item combined analysis of each tracer with the dark matter density field.
\end{itemize}
Clearly, in a single-tracer analysis the tracer with the lowest galaxy bias yields the highest information on the growth rate $f$. This is evident from Eq.~(\ref{F1}), where $b_\mathrm{g}$ enters in the denominator of the first term and the shot noise $\mathcal{E}$ is negligible if $\mathcal{E}/b_\mathrm{g}^2P\ll1$. Since $b_\mathrm{g_1}$ is very similar in all five cases, the Fisher information from a single tracer cannot be increased much by any particular choice of tracer.

On the contrary, a combination of two tracers can cancel out sampling variance and therefore considerably increase the available information on $f$ if the tracers are selected appropriately. As evident from the dashed red lines in Fig.~\ref{fig4}, the multitracer term in Eq.~(\ref{F_ff_exp}) gains importance over the single-tracer term as moving from the left to the right panel. Again, the two tracers obtained by a low mass cut yield the worst results, owing to the fact that the shot noise of both tracers is super-Poissonian in this selection. It is more optimal to impose a high mass cut in order to benefit from the sub-Poissonian shot noise level of the heaviest halos \cite{Hamaus2010}. However, the highest Fisher information content is obtained when correlating the two orthogonally weighted fields $\delta_+$ and $\delta_-$. The main reason for this is the large relative galaxy bias between the two fields and their relatively low shot noise level [see Table~\ref{tab1} and Eq.~(\ref{F_ff_exp})].

We have also explored the possibility of splitting the halo catalog into more than two mass bins, considering up to $N=10$ tracers. We find that the Fisher information from multiple tracers increases with the number of bins $N$ and approaches the result obtained with the two weighted fields $\delta_+$ and $\delta_-$ in the limit of high $N$.

Finally, adding the information from the dark matter density field to each one of the tracers increases the information content on $f$. In this case, sampling variance inherent in the density field $\delta$ is known and can thus be removed from the halo fields directly. According to Eq.~(\ref{F_ff_m}), the Fisher information is inversely proportional to the shot noise of the tracer, so the lowest stochasticity weight $w_-$ yields the best results.

\subsection{Multitracer fit}
So far we have investigated the Fisher information content on the growth rate using multiple tracers of the LSS. The question of how to actually constrain parameters of interest from a data set in the optimal way will be answered in this section. For this task we want to maximize the likelihood function from Eq.~(\ref{likelihood}), which is equivalent to minimizing its negative logarithm, the chi square
\begin{equation}
\chi^2\equiv\sum_{\mathbf{k}}\frac{1}{2}\dg\T(\mathbf{k},\mu)\C^{-1}\dg\pT(\mathbf{k},\mu) + \frac{1}{2}\ln\left(\det\C\right) \;. \label{chi2}
\end{equation}
Here, the covariance matrix $\C$ is given by the clustering model of Eq.~(\ref{Cov}) and we sum over all individual Fourier modes $\dg(\mathbf{k},\mu)$ from our halo catalog. When we add the dark matter density field as an observable, we use the model from Eq.~(\ref{Covm}) and $\dd=\left(\dm,\dg\right)$ as our data vector.

\begin{figure*}[!t]
\centering
\resizebox{\hsize}{!}{
\includegraphics[trim=0 0 0 0,clip]{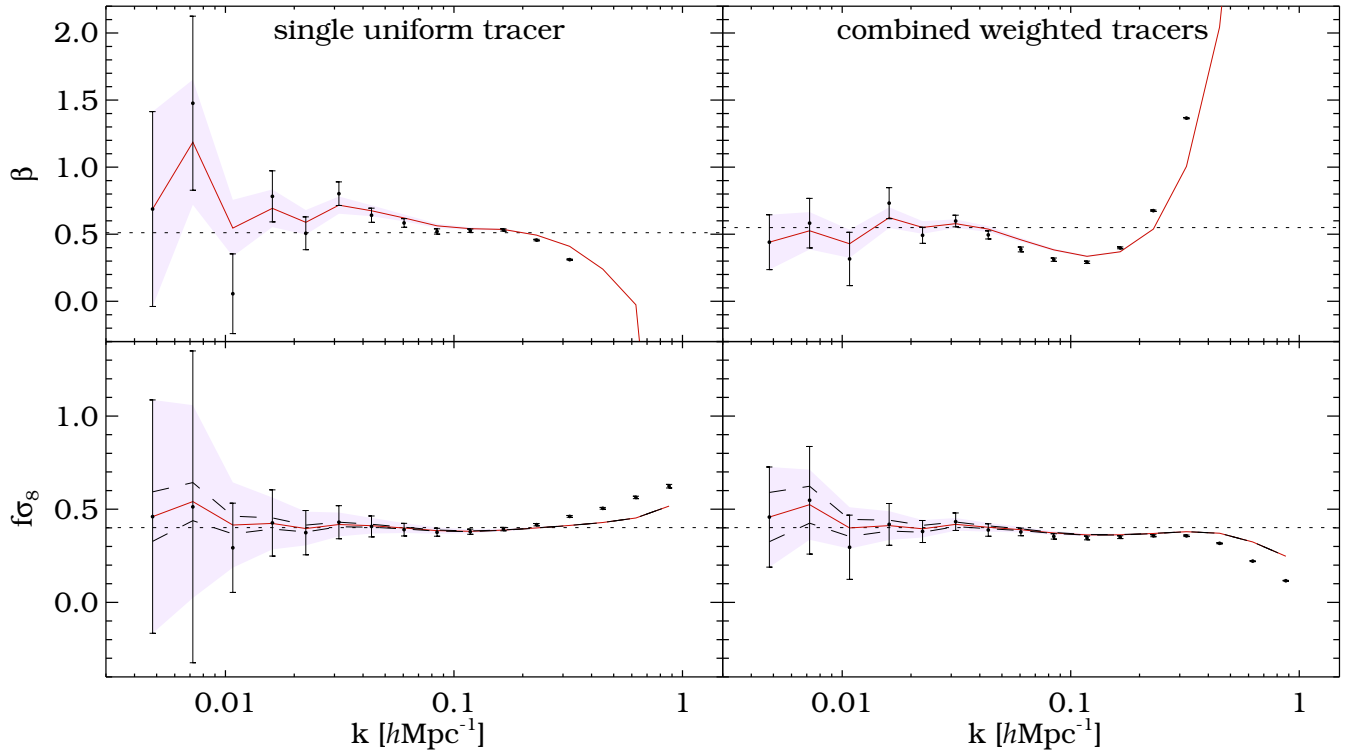}}
\caption{Fit for the redshift-space distortion parameter $\beta$ (top) and the product $f\sigma_8$ (bottom) from the two-point clustering statistics of halos in an $N$-body simulation with effective volume $V_\mathrm{eff}\simeq6.6h^{-3}\mathrm{Gpc}^3$ and halo-mass resolution $M_\mathrm{min}\simeq 9.4\times10^{11}\hMsun$ at $z=0$. LEFT: Conventional single-tracer analysis utilizing all halos (not weighted) from the same catalog. RIGHT: Multitracer analysis with the two fields $\delta_+$ and $\delta_-$, obtained through weighting the halo catalog with its principal components $w_+$ and $w_-$. The fitting results are shown in logarithmic bins of $k$ (points with 1-$\sigma$ error bars), as well as cumulative as a function of $k=k_\mathrm{max}$ with fixed $k_\mathrm{min}=0.0048h\mathrm{Mpc}^{-1}$ (red solid lines with shaded region). Dotted lines show the linear theory prediction with $f=\Om^{0.55}$ and $\sigma_8=0.81$, dashed lines the cumulative sampling variance limit.}
\label{fig5}
\end{figure*}

Figure~\ref{fig5} presents the fitting results for the RSD parameter $\beta$ and the product of growth rate $f$ with the normalization of the power spectrum $\sigma_8$ from our halo catalogs at redshift $z=0$. While $\beta$ has been obtained from a single-parameter fit, we have marginalized over the galaxy bias of each tracer as a free parameter in the fit for $f\sigma_8$. In the left column, the standard single-tracer analysis utilizing all objects in the halo catalog is performed. The best fits along with their $1\sigma$-error bars are shown both in $k$ bins (points with error bars), as well as cumulative as a function of $k=k_\mathrm{max}$ with fixed $k_\mathrm{min}=0.0048h\mathrm{Mpc}^{-1}$ (solid lines with shaded region).

The constraints on $\beta$ and $f\sigma_8$ are clearly affected by sampling variance, as evident from the large scatter of the points at low $k$. When the number of available Fourier modes grows towards higher $k$, this scatter becomes smaller; however, beyond a scale of $k\simeq0.2h\mathrm{Mpc}^{-1}$, linear theory breaks down and the fits depart from their scale-independent linear value assuming $f=\Om^{0.55}$ (dotted line). The cumulative sampling variance limit for the determination of $f\sigma_8$ is shown as a dashed line. Clearly, the single-tracer fit yields a substantially larger uncertainty compared to this limit.

\begin{figure*}[!t]
\centering
\resizebox{\hsize}{!}{
\includegraphics[trim=0 0 0 0,clip]{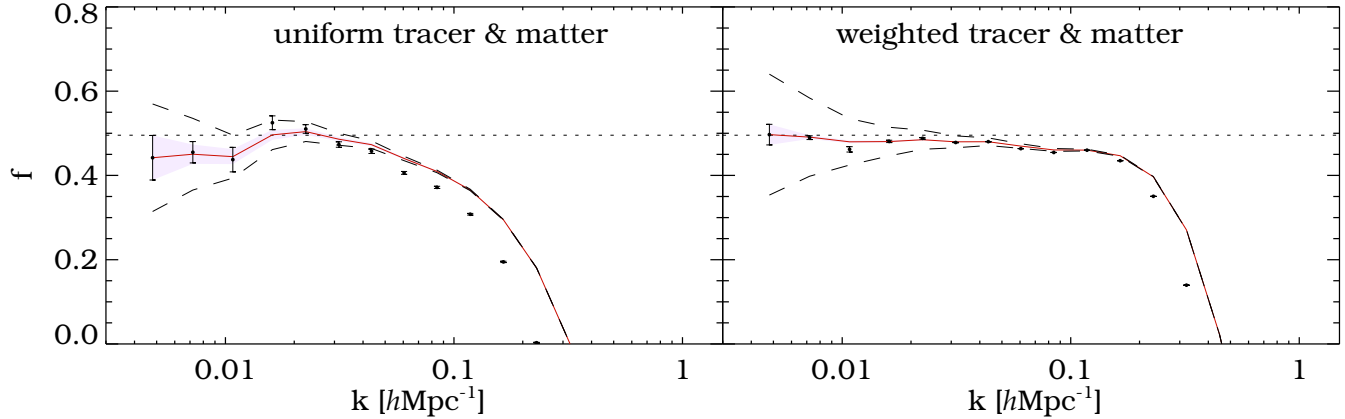}}
\caption{Fit for the growth rate $f$ from the two-point clustering statistics of halos \emph{and} the dark matter combined. LEFT: All uniform halos (not weighted) and the dark matter. RIGHT: All halos, weighted with the lowest stochasticity weight $w_-$, and the dark matter. The meaning of lines and symbols is the same as in Fig.~\ref{fig5}.}
\label{fig6}
\end{figure*}

When combining the two $w_+/w_-$-weighted tracers in a multitracer analysis as shown in the right column of Fig.~\ref{fig5}, the scatter of the fit is significantly suppressed. On the largest scales, the errors are reduced by up to a factor of 4 and the constraints on $f\sigma_8$ reach the sampling variance limit closely. However, the fit seems to deviate from the linear theory prediction already at $k\simeq0.04h\mathrm{Mpc}^{-1}$. This is likely due to the high bias of the $w_-$-weighted tracer: as shown in~\cite{Okumura2012}, more highly biased halos show a stronger scale dependence in redshift space, invalidating the Kaiser formula on even larger scales.
This could be corrected for by nonlinear RSD models, which is beyond the scope of this paper. On the other hand, the apparently more linear behavior of the single-tracer analysis may likely be coincidental at $k>0.04h\mathrm{Mpc}^{-1}$. This is supported by results shown in the left panel of Fig.~\ref{fig6}. Here, we combine the uniform halo catalog (without weighting) with the dark matter field from our simulation to fit for $f$ directly. Obviously, sampling variance has decreased even further, but deviations from linear theory already kick in at a scale of $k\simeq0.04h\mathrm{Mpc}^{-1}$.

Most impressive constraints on the growth rate are obtained when we combine the $w_-$-weighted halos with the dark matter density field, as depicted in the right panel of Fig.~\ref{fig6}. In this case, sampling variance has almost canceled out completely and the error bars on $f$ have diminished by up to a factor of 142 (at the peak of the power spectrum), when compared to the standard single-tracer analysis. Moreover, deviations from linear theory are very small up to scales of $k\sim0.2h\mathrm{Mpc}^{-1}$, making this kind of experiment the most promising one out of the four considered scenarios.

Methods to combine galaxy clustering and weak lensing data have been studied extensively in the recent literature (e.g., \cite{Baldauf2010,Gaztanaga2011,Cai2012,Cacciato2012,Simon2012,Jullo2012,VandenBosch2012}), suggesting high improvements for precision cosmology. Unfortunately, obtaining the dark matter density field in 3D from observations is a highly nontrivial problem and is subject of active research. Weak lensing tomography is the technique aiming to achieve this goal (e.g., \cite{Hu1999,Pen2004,Massey2007,Simon2009,VanderPlas2011}), but the resolution in the radial direction is not expected to be high because of the relatively broad lensing kernels along the line of sight. Moreover, we assume an ideal reconstruction of the dark matter density field without considering additional sources of error involved in the lensing measurement, such as \emph{shape noise} and \emph{intrinsic alignment}, for example. 

Without knowledge about $\delta$, in principle we would have to marginalize over all the parameters $\boldsymbol{\theta}^{(P)}$ of the dark matter power spectrum as well. This implies calculating the transfer function in each iteration of the fitting procedure, which goes beyond the scope of this paper. In the high signal-to-noise regime of the multitracer analysis, the degeneracy between the parameters $\boldsymbol{\theta}^{(P)}$ and $\boldsymbol{\theta}^{(b)}$ is expected to be rather weak (except the fundamental degeneracy between $f$ and $\sigma_8$), as the mixed terms between $\Sigma_i$ and $P_j$ in Eq.~(\ref{F_h}) are suppressed by $\Sigma$. Moreover, in Appendix~C of \cite{Hamaus2011} it has been shown that $P(k)$ cancels out to a high degree in the chi square of Eq.~(\ref{chi2}). Of course, in case the dark matter density field is available, we do not have to worry about those issues, since $b_\mathrm{g}$ and $P$ are directly observable.

\begin{figure*}[!t]
\centering
\resizebox{\hsize}{!}{
\includegraphics[trim=0 0 0 0,clip]{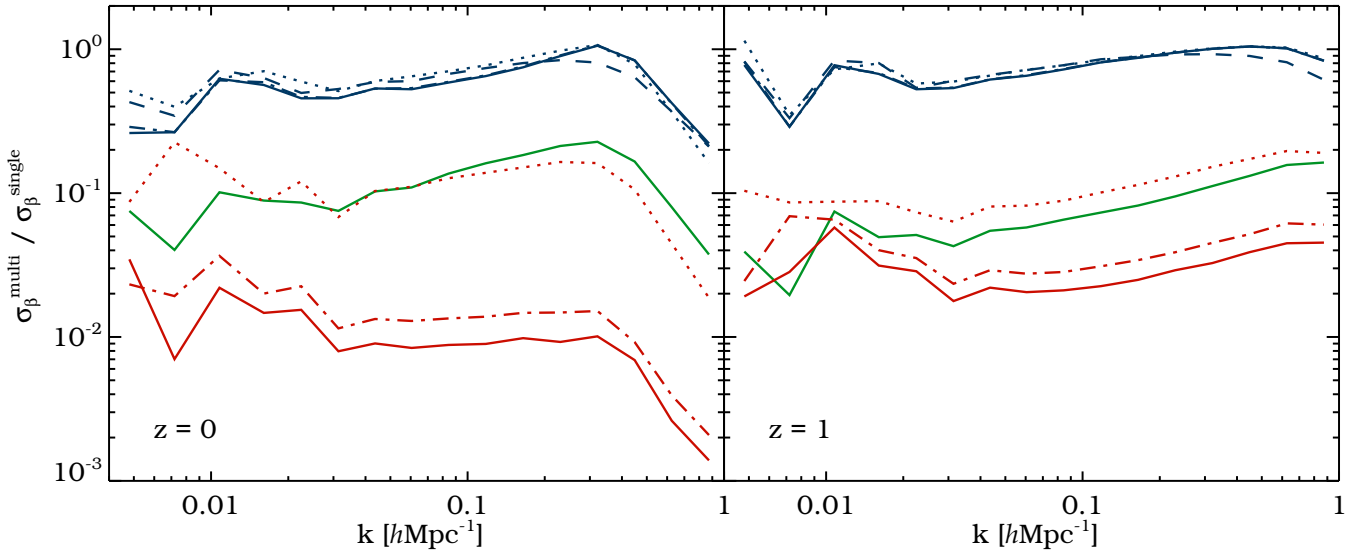}}
\caption{Ratios of the binned one-sigma error bars (from Fig.~\ref{fig5}) on $\beta$ (solid blue) and $f\sigma_8$ (dashed blue) when comparing the optimal multitracer to the single-tracer analysis at $z=0$ (left panel) and $z=1$ (right panel). Additionally, the improvement in constraints on $\beta$ from a combined clustering analysis of halos \emph{and} dark matter as compared to the single-tracer case is shown in solid green (no weighting of halos) and in solid red (optimally weighted halos). Adding a log-normal scatter of $\sigma_{\ln M}=0.1$ (dot-dashed) and $\sigma_{\ln M}=0.5$ (dotted) to the halo masses results in a degradation of the constraints from the weighted fields.}
\label{fig7}
\end{figure*}

In order to quantify the gains in accuracy, we compare the size of the error bars on $\beta$ and $f\sigma_8$ from the multitracer to the single-tracer analysis in Fig.~\ref{fig7} (solid and dashed blue lines, respectively). Obviously, sampling variance mostly cancels on the largest scales, yielding improvements in accuracy of up to a factor of 4. Beyond scales of $k\simeq0.1h\mathrm{Mpc}^{-1}$, the improvement is deteriorated due to the onset of nonlinear clustering and mode coupling. Deviations from linear theory increase towards smaller scales, making the fit of this model to the data increasingly biased (see Figs.~\ref{fig5} and \ref{fig6}).
Therefore, the drop of the curves at $k>0.3h\mathrm{Mpc}^{-1}$ is likely an artifact of the fitting procedure using an incorrect model and should not be trusted. At redshift $z=1$ this turnover is moved to smaller scales, the overall improvement in the error ratio is, however, slightly deteriorated. In order to access the cosmological information content buried in the semilinear regime of galaxy clustering in redshift space, one cannot avoid having to invoke more elaborate models involving perturbative methods, such as the ones proposed in~\cite{Seljak2011}.

More gains can be achieved when galaxies and ideal dark matter observations are combined, the improvement in the accuracy on $\beta$ compared to the ordinary single-tracer analysis amounts to about a factor of 10 in this case (solid green line). If, additionally, galaxies are weighted optimally, it increases by another factor of 10, two orders of magnitude better than what a single-tracer analysis can achieve. In this case the improvement even persists down to smaller scales of $k\simeq0.3h\mathrm{Mpc}^{-1}$. However, the effect of optimal weighting is diminished towards higher redshifts.

Unfortunately, the halo masses we used to construct our weighted density fields are not directly observable in reality. Yet, they correlate with many observables, such as X-ray luminosity, galaxy richness, weak lensing shear, velocity dispersion or the thermal Sunyaev-Zel'dovich (SZ) effect (e.g., \cite{Allen2011}). Scaling relations between these observables and halo mass can be calibrated with numerical simulations to obtain unbiased mass proxies with minimal scatter (e.g., \cite{Angulo2012,Noh2012}). While optical methods show a rather large scatter of $\sigma_{\ln M}\simeq0.45$ \cite{Rozo2009}, X-ray or SZ observations yield tighter relations with $\sigma_{\ln M}$ reaching below $0.1$ \cite{Kravtsov2006}.

We artificially add a constant log-normal scatter to the halo masses of our catalog in order to mimic the observational uncertainties in the mass determination for a rather pessimistic scenario ($\sigma_{\ln M}=0.5$) and a more optimistic scenario ($\sigma_{\ln M}=0.1$). The results are depicted in Fig.~\ref{fig7} as dotted and dot-dashed lines, respectively. While the constraints from the multitracer analysis on $\beta$ are only marginally affected for $\sigma_{\ln M}=0.1$ and degrade by roughly 20--30\% for $\sigma_{\ln M}=0.5$, the combined analysis using optimally weighted halos and the dark matter is more severely deteriorated by mass scatter. Here, even the optimistic scenario increases the uncertainty on the growth rate by roughly 50\%, while the benefits from optimal weighting are completely lost when going to $\sigma_{\ln M}=0.5$.

Clearly, the high level of precision that can be obtained with this kind of experiment demands precise mass estimates. Fortunately, a whole industry of existing and planned experiments devoted to cluster cosmology will provide those high quality data (e.g., \cite{ACT,DES,eROSITA,LSST,SPT,WFXT}).


\subsection{Halo model predictions \label{sec:HM}}
In this section we want to investigate how our results depend on the resolution of the simulation; respectively, the minimum resolved halo mass. In a real experiment, this corresponds to the depth of the galaxy survey with a corresponding luminosity threshold. Because $N$-body simulations with a given volume become increasingly expensive with higher resolution, we will turn to theoretical predictions henceforth. 

The clustering properties of dark matter and halos can be neatly described by the \emph{halo model} (see, e.g., \cite{Seljak2000}). The basic idea is to separately describe the clustering within a given halo (one-halo term) and the clustering amongst different halos (two-halo term). In \cite{Hamaus2010} the halo model is utilized to derive an analytical expression for the shot noise matrix. The result can be written as
\begin{equation}
\E=\bar{n}^{-1}\I-\bb\Mr\T-\Mr\bb\T \;, \label{E_hm}
\end{equation}
where $\bar{n}$ is the number density of halos per bin, $\I$ the identity matrix, $\Mr\equiv\M/\rhom-\bb\langle nM^2\rangle/2\rhom^2$ and $\M$ a vector containing the mean halo mass of each bin.
In Appendix~\ref{app_B} we utilize this expression to derive the clustering signal-to-noise ratio $\Sigma$ (as well as $\Sigma_i$ and $\Sigma_{ij}$) from the halo model. With Eq.~(\ref{F_ff}) we can then determine the Fisher information on the growth rate and compare the single-tracer analysis to the multitracer analysis.

\begin{figure}[!t]
\centering
\resizebox{\hsize}{!}{
\includegraphics[trim=0 0 0 0,clip]{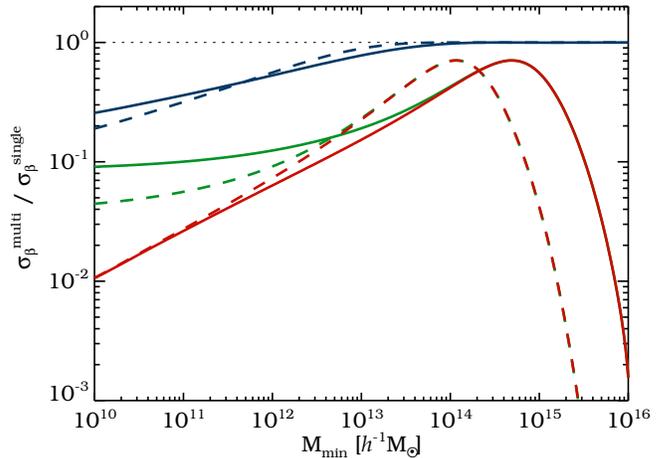}}
\caption{Halo model prediction for the improvement on $\sigma_\beta$ from the single-tracer analysis to the optimal multitracer analysis (blue) at $k\simeq0.016h\mathrm{Mpc}^{-1}$ as a function of the lower halo-mass threshold $M_\mathrm{min}$ at $z=0$ (solid) and $z=1$ (dashed). The results for the combined analysis of uniform halos (not weighted) and dark matter (green), as well as the combination of optimally weighted halos with the dark matter (red) are also depicted.}
\label{fig8}
\end{figure}

Figure~\ref{fig8} displays the same ratio as Fig.~\ref{fig7} for the uncertainty on $\beta$, but now as a function of minimum halo mass $M_\mathrm{min}$ at a fixed scale of $k\simeq0.016h\mathrm{Mpc}^{-1}$ (peak of the power spectrum). The gains from the multitracer method kick in at $M_\mathrm{min}\simeq10^{14}\hMsun$, where $\Sigma\simeq1$, and increase towards lower $M_\mathrm{min}$ due to the growing signal-to-noise ratio.

A combination of galaxy and dark matter observations may increase these gains further, especially when the galaxies are weighted optimally. In this case there is no saturation towards lower $M_\mathrm{min}$ and the improvement compared to the single-tracer analysis continues to grow. In contrast, the uniform galaxy overdensity field (not weighted) combined with dark matter already shows up a saturation at $M_\mathrm{min}\simeq10^{10}\hMsun$, so no more information on the growth rate can be gained when even more lighter halos are included in this kind of analysis. Also note that the halo model predictions underestimate the improvements obtained when adding dark matter clustering information, as our $N$-body results with $M_\mathrm{min}\simeq 9.4\times10^{11}\hMsun$ yield higher improvements (see Fig.~\ref{fig7}).

In the shot noise dominated regime above a minimum mass of $M_\mathrm{min}\sim10^{14}\hMsun$ the error ratio decreases again towards higher $M_\mathrm{min}$ because the Fisher information on $f$ from a single tracer here roughly scales as $\mathcal{E}^{-2}$, while for a tracer combined with dark matter as $\mathcal{E}^{-1}$ [see Eqs.~(\ref{F1}) and (\ref{F_m1}), respectively]. We refer the reader to Figs.~10 and 11 of \cite{Hamaus2011}, where a similar plot is shown for the individual error bars on the non-Gaussianity parameter $f_{\mathrm{NL}}$.

\section{Conclusions \label{sec3}}
In this paper we investigated the benefits of using weights in a multitracer analysis of LSS, with a particular focus on constraining the growth rate of structure formation. On the basis of earlier results on the clustering properties of dark matter halos and their stochasticity \cite{Hamaus2010}, we argue that the gains from a multitracer analysis in the sense of \cite{McDonald2009} can be achieved by considering only the two principal components of the clustering signal-to-noise ratio $\Sigma$ (or, equivalently, the two non-Poisson eigenvectors of the shot noise matrix $\E$).

We present their explicit functional forms in terms of weights, showing that the first one coincides with the weighting function explored in previous work \cite{Hamaus2010}, giving rise to low stochasticity and high bias. For the second one the weights are also mass dependent, but have a zero crossing, such that the overall bias is low. This yields a high relative galaxy bias $\alpha$ between the two tracers, maximizing the Fisher information content on the cosmological parameters \cite{McDonald2009}. All of the other eigenvectors oscillate around zero and add very little information. The advantage of reducing the information to two eigenvectors is that all of the objects in a given catalog can be used to construct the two principal components, while in the conventional multitracer analysis the catalog has to be split into many lower number density subsamples with higher shot noise \cite{Gil-Marin2010,Bernstein2011}. 

On the basis of numerical $N$-body simulations of dark matter halos, we demonstrate that the constraints on $\beta$ and $f\sigma_8$ can be improved by up to a factor of 4 relative to a single-tracer method, but most of the improvement comes from large scales (low $k$), while for higher $k$ the gains are smaller and vanish above $k\sim 0.1h\mathrm{Mpc}^{-1}$, where nonlinear effects introduce additional stochasticity between the two tracers. This technique is fairly insensitive to the observational uncertainty on the halo masses, as even a $50\%$ log-normal scatter does not degrade the improvements significantly. Halo model considerations suggest even higher gains of the method with increasing mass resolution.

One potential concern for our method is the possibility that galaxies might be bad tracers of their host-halo centers \cite{Skibba2011} and therefore exhibit less pronounced principal components in the clustering signal-to-noise ratio that are distinct from Poisson sampling. However, there are strong indications that certain types of galaxies do show strong correlations in both position and mass with their host halo, e.g. luminous red galaxies (LRGs) \cite{Zheng2009}. Techniques to distinguish satellite galaxies from central galaxies have been developed and the satellite fraction can be used as an estimator of the host-halo mass \cite{Zheng2005}.
Mock LRG catalogs obtained from a halo occupation distribution (HOD) model suggest some reduction in stochasticity is possible even without explicit knowledge of halo masses~\cite{Cai2011}. Therefore, an achievement of the presented gains seems feasible in light of upcoming spectroscopic galaxy surveys such as EUCLID~\cite{EUCLID}, which will attain galaxy number densities, host-halo mass ranges and a survey volume comparable to the simulations used in this paper~\cite{Laureijs2011,Majerotto2012}.

Whether these gains translate into a useful constraint on the final cosmological parameters depends on our ability to model nonlinear RSD effects. We find nonlinear effects are important for $\beta$ already at $k>0.03h\mathrm{Mpc}^{-1}$, although they appear to be important for $f \sigma_8$ only at $k>0.1h\mathrm{Mpc}^{-1}$. In the most pessimistic case where the RSD model cannot be trusted for $k>0.03h\mathrm{Mpc}^{-1}$, the multitracer method provides major gains relative to the single-tracer case, but neither method provides very strong constraints overall because of the limited number of available Fourier modes. In the case where we can use all the modes up to $k\sim 0.1h\mathrm{Mpc}^{-1}$, the overall errors are considerably smaller and the multitracer method provides less of an advantage. It is clear that a better modeling of the nonlinear effects in RSD is needed to understand the ultimate reach of RSD in both single-tracer and multitracer methods. 

In a more idealistic scenario, we also consider the joint analysis of halos and the dark matter density field, which in principle is achievable via a combination of spectroscopic redshift surveys and weak lensing tomography. Here, utilizing optimal weights can yield up to two orders of magnitude improvements in constraining $\beta$ as compared to a single-tracer analysis, but the method is more prone to uncertainties in the halo mass estimates. It is unlikely that this gain can be achieved in practice, since it is very difficult to measure dark matter clustering in the radial direction directly.

A further technique to construct differently biased tracers of the density field makes use of nonlinear transformations \cite{Seljak2012}. Although it is difficult to describe the effects of a nonlinear transformation on both signal and noise in galaxy clustering data, combined with optimal weights this may provide another tool for the multitracer analysis. 

In this paper we have focused on the information that can be extracted from RSD, in particular $\beta$ and $f\sigma_8$, but our method is not limited to constraints on the growth rate, but may be applied to the analysis of primordial non-Gaussianity \cite{Hamaus2011}, general relativistic corrections in large-scale clustering \cite{Yoo2012}, the Alcock-Paczy\'nski test \cite{McDonald2009} or any other quantity that influences the effective bias of tracers of the density field. It is possible that a better model of nonlinear RSD may yield a more efficient multitracer method, where the gains relative to the single-tracer analysis described here on large scales can be extended to smaller scales. We leave these directions for the future.

\begin{acknowledgments}
We thank Jaiyul Yoo, Jonathan Blazek, Tobias Baldauf and Zvonimir Vlah for fruitful discussions and Volker Springel for making public his $N$-body code {\scshape gadget}-2. This work is supported by the Packard Foundation, the Swiss National Foundation under Contract No. 200021-116696/1 and WCU Grant No. R32-10130. V.D. acknowledges support by the Swiss National Science Foundation. N.H. appreciated the hospitality of Lawrence Berkeley National Lab and the Institute for the Early Universe at Ewha University Seoul while parts of this work were completed.
\end{acknowledgments}


\appendix

\onecolumngrid
\vspace{2cm}

\section{FISHER MATRIX FOR MULTIPLE BIASED TRACERS \label{app_A}}
With Eqs.~(\ref{C'}) and (\ref{Sherman-Morrison}) plugged into Eq.~(\ref{fisher}), we have
\begin{multline}
F_{ij}=\frac{1}{2}\mathrm{Tr}\left[\left(\bg\bg_i\T+\bg_i\bg\T+\bg\bg\T\frac{P_i}{P}\right)\C^{-1}\left(\bg\bg_j\T+\bg_j\bg\T+\bg\bg\T\frac{P_j}{P}\right)\C^{-1}\right]P^2= \\
=\frac{1}{2}\left(
\bg\T\C^{-1}\bg\bg_i\T\C^{-1}\bg_j + \bg\T\C^{-1}\bg\bg_j\T\C^{-1}\bg_i + \bg\T\C^{-1}\bg_i\bg\T\C^{-1}\bg_j + \bg_i\T\C^{-1}\bg\bg_j\T\C^{-1}\bg + 
\bg\T\C^{-1}\bg\bg\T\C^{-1}\bg_i\frac{P_j}{P} +\right. \\ \left. \bg\T\C^{-1}\bg\bg_i\T\C^{-1}\bg\frac{P_j}{P} +
\bg\T\C^{-1}\bg\bg\T\C^{-1}\bg_j\frac{P_i}{P} + \bg\T\C^{-1}\bg\bg_j\T\C^{-1}\bg\frac{P_i}{P} +
\bg\T\C^{-1}\bg\bg\T\C^{-1}\bg\frac{P_iP_j}{P^2}
\right)P^2= \\
=\frac{\Sigma}{1+\Sigma}\left[\Sigma_{ij}-\frac{\Sigma_i\Sigma_j}{1+\Sigma}\right] + 
\frac{\Sigma_i\Sigma_j}{\left(1+\Sigma\right)^2} + 
\frac{\Sigma_i\Sigma}{\left(1+\Sigma\right)^2}\frac{P_j}{P} + 
\frac{\Sigma_j\Sigma}{\left(1+\Sigma\right)^2}\frac{P_i}{P} +
\left(\frac{\Sigma}{1+\Sigma}\right)^2\frac{P_iP_j}{2P^2}= \\
=\left[\Sigma^{-1}\left(\Sigma_{ij}+\frac{\Sigma_i\Sigma_j}{\Sigma}\right) +
\left(\Sigma_{ij}-\frac{\Sigma_i\Sigma_j}{\Sigma}\right) +
\frac{\Sigma_i}{\Sigma}\frac{P_j}{P} + 
\frac{\Sigma_j}{\Sigma}\frac{P_i}{P} +
\frac{P_iP_j}{2P^2}\right]\left(1+\Sigma^{-1}\right)^{-2}\;. \label{F}
\end{multline}
With additional knowledge about the dark matter density field we need to work out Eq.~(\ref{fisher}) by plugging in Eqs.~(\ref{Covm'}) and (\ref{CovmI}). Let us first note that
\begin{equation}
\frac{\partial\C}{\partial\theta_i}\C^{-1} =
\left( \begin{array}{cc}
P_i/P-\Sigma_i & \bg_i\T\E^{-1}P \\
\bg_i+\bg\left(P_i/P-\Sigma_i\right)\;\; & \;\;\bg\bg_i\T\E^{-1}P \\
\end{array} \right)  \;, \nonumber
\end{equation}
so
\begin{gather}
\frac{\partial\C}{\partial\theta_i}\C^{-1}\frac{\partial\C}{\partial\theta_j}\C^{-1} =\nonumber \\
\left( \begin{array}{cc}
\left(P_i/P-\Sigma_i\right)\left(P_j/P-\Sigma_j\right)+\Sigma_{ij}+\Sigma_i\left(P_j/P-\Sigma_j\right) & \left(P_i/P-\Sigma_i\right)\bg_j\T\E^{-1}P+\Sigma_i \bg_j\T\E^{-1}P\\
\left[\bg_i+\bg\left(P_i/P-\Sigma_i\right)\right]\left(P_j/P-\Sigma_j\right)+\bg\Sigma_{ij}+\bg\Sigma_i\left(P_j/P-\Sigma_j\right)\;\; & \;\;\bg_i\bg_j\T\E^{-1}P+\bg\bg_j\T\E^{-1}P\left(P_i/P-\Sigma_i\right)+ \bg\Sigma_i\bg_j\T\E^{-1} P \\
\end{array} \right) \nonumber
\end{gather}
This yields
\begin{equation}
F_{ij}=\frac{1}{2}\mathrm{Tr}\left(\frac{\partial\C}{\partial\theta_i}\C^{-1}\frac{\partial\C}{\partial\theta_j}\C^{-1}\right)=\Sigma_{ij}+\frac{P_iP_j}{2P^2} \;.
\end{equation}

\section{HALO MODEL PREDICTION FOR THE CLUSTERING SIGNAL-TO-NOISE RATIO \label{app_B}}
\noindent
In the halo model the shot noise matrix is given by Eq.~(\ref{E_hm}). In order to invert $\E$, we write $\E=\A-\Mr\bb\T$ with $\A\equiv\bar{n}^{-1}\I-\bb\Mr\T$ and apply the Sherman-Morrison formula twice:
\begin{equation}
\E^{-1}=\A^{-1}+\frac{\A^{-1}\Mr\bb\T\A^{-1}}{1-\bb\T\A^{-1}\Mr} \;\;\;\;,\;\;\;\; \A^{-1}=\bar{n}\I+\frac{\bb\Mr\T\bar{n}}{\bar{n}^{-1}-\Mr\T\bb} \;.
\end{equation}
With
\begin{gather}
\Sigma_{ij}\equiv\bg_i\T\E^{-1}\bg_jP=\frac{\bg_i\T\A^{-1}\bg_j\left(1-\bb\T\A^{-1}\Mr\right)+\bg_i\T\A^{-1}\Mr\bb\T\A^{-1}\bg_j}{1-\bb\T\A^{-1}\Mr}P \;,
\end{gather}
and
\begin{gather}
\bg_i\T\A^{-1}\bg_j=\frac{\bg_i\T\bg_j\left(\bar{n}^{-1}-\Mr\T\bb\right)+\bg_i\T\bb\Mr\T\bg_j}{\bar{n}^{-1}-\Mr\T\bb}\bar{n} \;, \\
1-\bb\T\A^{-1}\Mr=\frac{\left(\bar{n}^{-1}-\Mr\T\bb\right)^2-\bb\T\bb\Mr\T\Mr}{\bar{n}^{-1}-\Mr\T\bb}\bar{n} \;,
\end{gather}
and similar terms combining $\bg$, $\bg_i$, $\bb$ and $\Mr$, after some algebra we get
\begin{gather}
\Sigma_{ij}=\bg_i\T\bg_j\bar{n}P+\frac{\bg_i\T\Mr\bb\T\bb\Mr\T\bg_j+\left(\bg_i\T\bb\Mr\T\bg_j+\bg_i\T\Mr\bb\T\bg_j\right)\left(\bar{n}^{-1}-\Mr\T\bb\right)+\bg_i\T\bb\Mr\T\Mr\bb\T\bg_j}{\left(\bar{n}^{-1}-\Mr\T\bb\right)^2-\bb\T\bb\Mr\T\Mr}\bar{n}P \;.
\end{gather}
In the continuous limit ($N\rightarrow\infty$), we can exchange the vector products by integrals over the mass function and set $\bar{n}_\mathrm{tot}=\bar{n}N$. This finally yields
\begin{equation}
\Sigma_{ij}=\langle b_ib_j\rangle\bar{n}_{\mathrm{tot}}P+\frac{\langle b_\mathrm{g}^2\rangle\langle\mathcal{M}b_i\rangle\langle\mathcal{M}b_j\rangle+\left(\langle b_ib_\mathrm{g}\rangle\langle \mathcal{M}b_j\rangle+\langle\mathcal{M}b_i\rangle\langle b_jb_\mathrm{g}\rangle\right)\left(\bar{n}_{\mathrm{tot}}^{-1}-\langle \mathcal{M}b_\mathrm{g}\rangle\right)+\langle b_ib_\mathrm{g}\rangle\langle b_jb_\mathrm{g}\rangle\langle\mathcal{M}^2\rangle}{\left(\bar{n}_{\mathrm{tot}}^{-1}-\langle\mathcal{M}b_\mathrm{g}\rangle\right)^2-\langle b_\mathrm{g}^2\rangle\langle\mathcal{M}^2\rangle}\bar{n}_{\mathrm{tot}}P \;.
\end{equation}

\vspace{0.6cm}
\twocolumngrid


\end{document}